\documentstyle[aps,prb]{revtex}
\input epsf.sty

\begin{document}

\draft
\wideabs{
\title{Near field and far field scattering of surface plasmon polaritons
by one-dimensional surface defects}

\author{J. A. S{\'a}nchez-Gil}
\address{Instituto de Estructura de la Materia, Consejo Superior de
Investigaciones Cient{\'\i}ficas \\
Serrano, 121, 28006 Madrid, Spain}

\author{A. A. Maradudin}
\address{Department of Physics and Astronomy, and Institute
for Surface and Interface Science \\ University of California,
Irvine, CA 92697}
\date{April 7, 1999}
\maketitle
\begin{abstract}

A rigorous formulation for the scattering of surface plasmon polaritons
(SPP) from a one-dimensional surface defect of any shape that yields the
electromagnetic field in the vacuum half-space above the vacuum-metal
interface is developed by the use of an impedance boundary condition.
The electric and magnetic near fields, the angular
distribution of the far-field radiation into vacuum due to SPP-photon
coupling, and the SPP reflection and transmission coefficients are
calculated by numerically solving the $k$-space integral equation upon
which the formulation is based.  In particular, we consider
Gaussian-shaped defects (either protuberances or indentations) and study
the dependence of the above mentioned physical quantities on their $1/e$
half-width $a$ and height $h$. SPP reflection is significant for narrow
defects ($a\stackrel{<}{\sim} \lambda/5$, for either protuberances or
indentations, where $\lambda$ is the wavelength of the SPP); maximum
reflection ({\it plasmon mirrors}) is achieved for $a\approx\lambda/10$.
For increasing defect widths, protuberances and
indentations behave differently. The former give rise to a monotonic
increase of radiation at the expense of SPP transmission for increasing
defect half-width. However, indentations exhibit a significant increase
of radiation (decrease of SPP transmission) for half-widths of the order
of or smaller than the wavelength, but tend to total SPP transmission
in an oscillatory manner upon further increasing the half-width. Both the
position of the maximum radiation  and the oscillation period depend on
the defect height, which in all other cases  only affects the process
quantitatively. {\it Light-emitters} might thus be associated with either
wide indentations, or protuberances  with widths that are of the order of
or smaller than the wavelength.

\end{abstract}
}

\section{INTRODUCTION}
\label{sec_int}

In this paper, we study the scattering of surface plasmon polaritons
(SPP) by surface defects. SPP are $p$-polarized electromagnetic (EM)
waves bound to a dielectric-metal interface and caused by the surface
oscillations of the electron plasma of the metal.\cite{sp} They
propagate along the metal interface a distance of the order of the SPP
mean free path (ranging from microns in the visible to millimeters in
the infrared, of course depending also on the metal being considered),
undergoing scattering processes due to surface roughness. This
constitutes a classical problem of fundamental interest not only in the
case of individual defects (cf. Ref. \onlinecite{pmbg94} and references
therein), but also for periodically or randomly (or both)
distributed defects.\cite{pg96,prb96,kits96,prb97} Furthermore,
it is obviously crucial in any light scattering problem involving rough
metal surfaces where roughness-induced excitation of SPP occurs. This
has been explicitly shown in connection with either single defects
\cite{psg94,chema95,alb} or random corrugation,\cite{psg95,wo,mmm95,owm98}
the latter configuration being relevant to
the phenomenon of (SPP mediated) enhanced backscattering of light.
In addition to that, light-SPP coupling plays a central role in other
phenomena such as anomalous transmission through metal slabs with hole
arrays,\cite{ebbe98,schro98} surface-enhanced Raman scattering,
\cite{sers,vp96,jcp98} or biosensing.\cite{bio}

In recent years, the advent of near-field optical microscopy \cite{nf}
has opened up the possibility to study experimentally SPP in a direct
manner. Among the various configurations developed, photon scanning
tunneling microscopy \cite{pstm} (PSTM), basically exploiting SPP
excitation in the attenuated total reflection arrangement, has made it
possible to probe the SPP structure,\cite{marti,adam} localized SPP on
randomly rough surfaces,\cite{bozh95} and SPP resonances in fractal
colloid clusters \cite{tsai94} and single particles.\cite{nie97,klar98}
Moreover, PSTM images have been obtained by surface-enhanced Raman
scattering probing single molecules adsorbed on single nanoparticles.
\cite{nie97} PSTM in combination with direct-write lithography has
made it possible to create sub-micron defects on metal
surfaces.\cite{smol95}

Particularly relevant to the present work are the recent experimental
studies on SPP scattering by surface defects.\cite{smol,bozh,smol99}
These studies have shown evidence of drastically distinct scattering
properties depending on the defect size. Specifically, surface defects
favoring SPP reflection and light coupling, called SPP mirrors and
flashlights,\cite{smol} respectively, have been described, as well as
SPP microlenses and microcavities;\cite{bozh} SPP Bloch waves have also
been imaged in periodic arrays of surface defects.\cite{smol99}
Interestingly, the possibilities of artificially creating micro-optical
components for SPP have also been noted in these studies. Much has to be
done, however, from the theoretical standpoint. Quite recently, calculations
for circularly symmetric defects have successfully accounted for the
peculiar azimuthal dependence of the radiated pattern;\cite{snm97} in
addition, such calculations have been used to retrieve the surface profile.
\cite{pasc98} In the case of one-dimensional surface defects, preliminary
calculations have focused on the optimization of the defect size to obtain
SPP mirrors and so-called light-emitters.\cite{apl98} In this regard, it
is our purpose to address in detail the SPP scattering by one-dimensional
surface defects, including near-field and far-field calculations (along
with energy balance) and their dependence on defect size parameters. Thus
we expect not only to shed light on the experimental works mentioned above,
but also to find and predict related effects.

The physical system we consider here is a planar one-dimensional
metal surface with a one-dimensional defect. The surface corrugation
is modeled by using a local impedance boundary condition (IBC) on a flat
surface. The connection between surface impedance and real surface
corrugation has been recently demonstrated,\cite{aamibc} and its
validity to give accurate quantitative results has been shown
in numerical calculations of grating-induced SPP-photon coupling.
\cite{prb96} A scattering-theoretic formulation of the interaction of
an SPP with the surface roughness is developed by imposing the IBC
on the amplitude of the magnetic field in the vacuum region in the
form of a Rayleigh expansion. Upon solving the resulting integral
equation for the scattering amplitude, the magnetic field at any
point in the vacuum half-space can be calculated. We will focus on
the far field angular distribution and the surface field amplitudes
to determine, respectively, the total radiated energy $S$, and the
SPP reflection $R_{SP}$ and transmission $T_{SP}$ coefficients. By
numerical simulation calculations, these quantities are computed.

The paper is organized as follows. The theoretical formulation
is derived in Sec.~\ref{sec_the}, and some details pertaining to the
numerical procedure are given in the Appendix. In Sec.~\ref{sec_res}, we
show the results obtained for a single Gaussian defect and the influence
of defect width and height. Finally, Section~\ref{sec_con} summarizes the
conclusions drawn from this research.

\section{THEORY}
\label{sec_the}

\subsection{Scattering Equations}
\label{sec_the_se}

\begin{figure}
\epsfxsize=2.5in \epsfbox{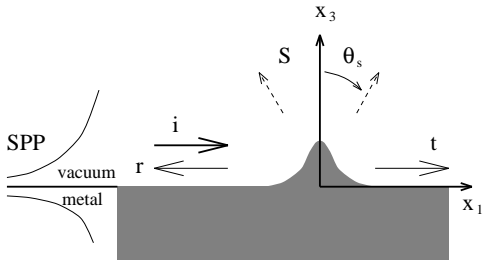}
\caption{Illustration of the scattering geometry.}
\label{fig_plasmon}
\end{figure}

We  study the scattering of a p-polarized SPP of frequency
$\omega$ propagating along a flat vacuum-metal interface ($x_3=0$)
by a one-dimensional obstacle (constant along the $x_2$-axis, see
Fig.~\ref{fig_plasmon}). Under these circumstances, the three-dimensional
electromagnetic problem can be cast into a two-dimensional scalar
problem in such a way that the single, nonzero component of the
magnetic field amplitude $H_2(x_1,x_3)$ is the solution of the
corresponding two-dimensional Helmholtz equation in the upper
(vacuum) and lower (metal) half spaces. The magnetic field in vacuum
is assumed to be the sum of an incoming SPP and a scattered field as
follows
\begin{eqnarray}
 \lefteqn{H_2^>(x_1,x_3)  =\exp[ik(\omega)x_1-\beta_o(\omega)x_3]} &&
    \nonumber \\ && \phantom{+++}+\int_{-\infty}^{\infty}\frac{dq}{2\pi}
    \, R(q,\omega)\exp[iqx_1+i\alpha_o(q,\omega)x_3] ,
\label{eq_h}
\end{eqnarray}
where
\begin{eqnarray}
    k(\omega) & \equiv k^R(\omega)+ik^I(\omega)={\displaystyle
      \frac{\omega}{c}\left(1-\frac{1}{\epsilon(\omega)}
      \right)^{1/2},}
\label{eq_kw}
\end{eqnarray}
\begin{eqnarray}
    \beta_0(\omega) & = & \left(k(\omega)^2-\frac{\omega^2}{c^2}
      \right)^{1/2}=\frac{\omega}{c}[-\epsilon(\omega
      )]^{-1/2},
\label{eq_bw}
\end{eqnarray}
and
\begin{mathletters}\begin{eqnarray}
    \alpha_o(q,\omega) & = & \left(\frac{\omega^2}{c^2}-q^2
      \right)^{1/2}\;\;\;\; |q|\leq\frac{\omega}{c} \\
     & = & i\left(q^2-\frac{\omega^2}{c^2}\right)^{1/2}
      \;\;\; |q|>\frac{\omega}{c}.
\end{eqnarray}\end{mathletters}
Note that the expressions for the  SPP wavevector components $k(\omega)$
and $\beta_0(\omega)$ in vacuum apply in the limit $|\epsilon(\omega)|\gg
1$. This stems from the fact that the continuity conditions
across the interface are mapped onto a local IBC on the planar
surface $x_3=0$ in the form
\begin{eqnarray}
 \lefteqn{\left.\frac{\partial}{\partial x_3} H_2^>(x_1,x_3)\right|_{x_3=0}}
       && \nonumber \\ &&
  =-\frac{\omega}{c}\,\frac{1+s(x_1)}{[-\epsilon (\omega)]^{ 1/2}}
      \;H_2^>(x_1,x_3)|_{x_3=0} ,
\label{eq_ibc}
\end{eqnarray}
where the superscript $>$ indicates the vacuum region,
$-(\omega/c)[-\epsilon (\omega )]^{-1/2}s(x_1)$ is the contribution to the
surface impedance associated with the obstacle, and $\epsilon(\omega)$
is the isotropic, frequency-dependent dielectric function of the metal.
The IBC has been widely used in the past to  model the vacuum-metal
interface qualitatively, especially in the infrared region of the optical
spectrum. Furthermore, it has been recently proven to be quantitatively
accurate in calculations of grating-induced photon-SPP coupling \cite{prb96}
by using the connection between surface impedance and real corrugation
demonstrated in Ref. \onlinecite{aamibc}.

In order to calculate the scattering amplitude $R(q,\omega)$, we
substitute Eq. (\ref{eq_h}) into Eq. (\ref{eq_ibc}), and  obtain the
 integral equation $R(q,\omega )$ satisfies,
\begin{eqnarray}
    R(q,\omega)=&& G_0(q,\omega)V\bbox(q|k(\omega)\bbox) \nonumber\\ &&
     +G_0(q,\omega)\int_{-\infty}^{\infty}\! \frac{dp}{2\pi}
     V(q|p)R(p,\omega),
\label{eq_ierq}
\end{eqnarray}
where
\begin{eqnarray}
    G_0(q,\omega) & \equiv &
     \frac{i\epsilon(\omega)}{\epsilon(\omega)\alpha_0(q,\omega)
     +i(\omega/c)[-\epsilon(\omega)]^{1/2}}
\label{eq_gq}
\end{eqnarray}
is the Green's function of the SPP on the unperturbed surface
[$s(x_1)=0$]. We have also introduced the scattering potential
\begin{eqnarray}
    V(q|p) & \equiv & \beta_0(\omega)\hat{s}(q-p),
\label{eq_vq}
\end{eqnarray}
with
\begin{eqnarray}
    \hat{s}(Q)&=&\int_{-\infty}^{\infty}\! dx_1\; e^{-iQx_1}s(x_1),
\label{eq_sq}
\end{eqnarray}
to simplify the notation. Equation (\ref{eq_ierq}) can be rewritten in a
more convenient manner by substituting
\begin{eqnarray}
   R(q,\omega) &=& G_0(q,\omega)T(q,\omega),
\label{eq_rtq}
\end{eqnarray}
into it, so that
\begin{eqnarray}
    T(q,\omega)=&& V\bbox(q|k(\omega)\bbox) \nonumber \\ &&
    +\int_{-\infty}^{\infty}\!
    \frac{dp}{2\pi}V(q|p)G_0(p,\omega)T(p,\omega).
\label{eq_ietq}
\end{eqnarray}
Equation (\ref{eq_ietq}), along with Eqs. (\ref{eq_h}) and
(\ref{eq_rtq}), is the basis of our theoretical formulation.

In solving
Eq. (\ref{eq_ietq}), it is very important how we deal with the poles
appearing in the Green's function (\ref{eq_gq}). First, we rewrite the
latter in the form
\begin{eqnarray}
   G_0(q,\omega) & = & C(q,\omega)\left(
     \frac{1}{q-k(\omega)}-\frac{1}{q+k(\omega)}\right),
\end{eqnarray}
with
\begin{eqnarray}
   C(q,\omega) & \equiv & \frac{\epsilon(\omega)\alpha_0(q,\omega)-
   i(\omega/c)[-\epsilon(\omega)]^{1/2}}{2i\epsilon(\omega) k(\omega)}.
\end{eqnarray}
We now assume that the metal dielectric function is given by
Drude's expression
\begin{eqnarray}
    \epsilon(\omega) &=& 1-\frac{\omega_p^2}{\omega^2} ,
\end{eqnarray}
where $\omega_p$ is the plasma frequency, in the absence of
absorption losses.
Therefore, in light of Eq. (\ref{eq_kw}), we have to take the limit
$k^I(\omega)\rightarrow 0$ in Eq. (\ref{eq_gq}), to obtain
\begin{eqnarray}
    G_0(q,\omega) &&= C(q,\omega)\left(
     \left.\frac{1}{q-k^R(\omega)}\right|_P-\left.
     \frac{1}{q+k^R(\omega)}\right|_P\right. \nonumber \\ &&
     \left.\phantom{\frac{1}{2}}+\pi i[\delta\bbox(q-k^R(\omega)\bbox)
     +\delta\bbox(q+k^R(\omega)\bbox)]\right) .
\label{eq_gqri}
\end{eqnarray}
The first two terms on the right-hand side of Eq. (\ref{eq_gqri}) have
meanings in the Cauchy's principal value sense, whereas the last
two terms are delta functions. Once Eq. (\ref{eq_ietq}) is solved for
$T(q,\omega)$ [we will see below how to do so numerically with
the help of Eq. (\ref{eq_gqri})], we proceed to calculate the
electric and magnetic near fields, the SPP-photon coupling, and the
SPP reflection and transmission coefficients in the following manner.

\subsection{Near Field}
\label{sec_the_nf}

The magnetic field at any point in the vacuum half-space can be
straightforwardly calculated from Eq. (\ref{eq_h}), upon recalling
Eq. (\ref{eq_rtq}), which relates  $T(q,\omega)$ with the scattering
amplitude $R(q,\omega)$. Then the electric field components in vacuum are
easily written also as  functions of $R(q,\omega)$ by means of a Maxwell
curl equation as follows:
\begin{mathletters}\label{eq_e}\begin{eqnarray}
  E_1^>(x_1,x_3) = &&\frac{c}{\omega}i\beta_0(\omega)
      \exp[ik(\omega)x_1-\beta_0(\omega)x_3] \nonumber \\ &&
     + \frac{c}{\omega}\int_{-\infty}^{\infty}\frac{dq}{2\pi}
     \,\alpha_0(q,\omega) R(q,\omega) \nonumber \\ && \times
     \exp[iqx_1+i\alpha_0(q,\omega)x_3] \\
  E_2^>(x_1,x_3) = && 0 \\
  E_3^>(x_1,x_3) = &&-\frac{c}{\omega}k(\omega)
      \exp[ik(\omega)x_1-\beta_0(\omega)x_3] \nonumber \\ &&
     - \frac{c}{\omega}\int_{-\infty}^{\infty}
     \frac{dq}{2\pi}\, qR(q,\omega) \nonumber \\ && \times
     \exp[iqx_1+i\alpha_0(q,\omega)x_3].
\end{eqnarray}\end{mathletters}
The time-averaged Poynting vector thus reads:
\begin{eqnarray}
 <{\bf S}> = \frac{c}{8\pi} \Re \left({\bf E}\times{\bf H}^*\right) =
             \frac{c}{8\pi} \Re \left(-E_3 H_2^*,0,E_1 H_2^*\right) ,
\label{eq_pv}
\end{eqnarray}
where $\Re$ denotes the real part and the asterisk the complex
conjugate.

\subsection{Radiated energy}
\label{sec_the_s}

The total power carried away from the surface in the form of
volume electromagnetic waves propagating in the vacuum region
above it, per unit length of the system along the $x_2$-axis is
\begin{eqnarray}
  P_{sc}= & & \int_{-\infty}^{\infty}\! dx_1\; <S^{(sc)}_3> \nonumber \\
    = &&\frac{c^2}{8\pi\omega}
    \int_{-\omega/c}^{\omega/c}
    \frac{dq}{2\pi}\alpha_0(q,\omega)\mid\!R(q,\omega)\!\mid^2 .
\label{eq_psc}
\end{eqnarray}
Note that only the scattered field contribution to the $x_3$-component of
the time-averaged Poynting vector is used. Equation (\ref{eq_psc}) must
be normalized by the power carried by the incident SPP
per unit length along the $x_2$-axis
\begin{eqnarray}
  P_{inc}= & & \int_0^{\infty}\! dx_3\; <S^{(inc)}_1>
    =\frac{c^2k(\omega)}{16\pi\omega\beta_0(\omega)} ,
\label{eq_pin}
\end{eqnarray}
where $<S^{(inc)}_1>$ is the $x_1$-component of the time-averaged Poynting
vector of the incident SPP. Then the total, normalized scattered power
$S$ is given by
\begin{eqnarray}
    S &= {\displaystyle\frac{P_{sc}}{P_{inc}}} &
=\int_{-\pi/2}^{\pi/2}
      \! d\theta_s \frac{\partial R}{\partial\theta_s},
\label{eq_s}
\end{eqnarray}
where
\begin{eqnarray}
   \frac{\partial R}{\partial\theta_s} =& &{\displaystyle\frac{1}{2
    \pi}\frac{\beta_0(\omega)}{2k(\omega)}\;\alpha_0^2\left(
     q=\frac{\omega}{c}\sin\theta_s\right)} \nonumber\\ && \times\left|
     R\left(q=\frac{\omega}{c}\sin\theta_s\right)\right|^2 ,
\label{eq_drc}
\end{eqnarray}
is the differential reflection coefficient (DRC), namely, the fraction
of the energy of the incident SPP that is scattered into an angular
region of width $d\theta_s$ about the scattering direction $\theta_s$
where the scattering angle $\theta_s$ is measured clockwise with respect
to the $x_3$-axis (see Fig.~\ref{fig_plasmon}).

\subsection{Reflection and transmission coefficients}
\label{sec_the_rt}

In order to evaluate the amplitude of the reflected and transmitted SPP,
we study the behavior of $H_2^>(x_1,x_3)$, Eq. (\ref{eq_h}), with the help
of Eqs. (10) and (15), on the surface $x_3=0$. At this point, great care
has to be taken when calculating  the contribution to the scattered field
in Eq. (2) from the Cauchy principal value integrals arising from the first
two terms on the right-hand side of Eq. (15). We assume that the
obstacle has a finite extent and is centered about $x_1=0$. If
we focus on the regions $x_1\ll 0$ and $x_1\gg 0$ far from the
obstacle, it can be shown, by working out the contributions from
those integrals in the complex $q$-space with the help of Cauchy's
theorem,\cite{cauchy} that the magnetic field is given by:
\begin{mathletters}\label{eq_hxinf}\begin{eqnarray}
    H_2^>(x_1,x_3=0) =& &\exp[ik^R(\omega)x_1] \nonumber \\ & &
       +r(\omega)\exp[-ik^R(\omega)x_1], x_1\ll 0 \\
     = && t(\omega)\exp[ik^R(\omega)x_1], \phantom{+-} x_1\gg 0,
\end{eqnarray}\end{mathletters}
where the amplitudes of the reflected and transmitted SPP,
$r(\omega)$ and $t(\omega)$, respectively, are
\begin{mathletters}\label{eq_spprt}\begin{eqnarray}
  r(\omega) &=& iT\bbox(-k^R(\omega),\omega\bbox)\; C\bbox(-k^R(\omega),
    \omega\bbox) \\
  t(\omega) &=& 1+iT\bbox(k^R(\omega),\omega\bbox)\; C\bbox(k^R(\omega),
    \omega\bbox) .
\end{eqnarray}\end{mathletters}
Equations (\ref{eq_hxinf}) manifest the fact that, away from the
obstacle, only the incident and reflected SPP (on the left-hand side,
see Fig. 1) and the transmitted SPP (on the right-hand side)
propagate along the interface. The corresponding reflection
and transmission coefficients are
\begin{mathletters}\label{eq_spprt2}\begin{eqnarray}
   R(\omega) &=& \mid\! r(\omega)\!\mid^2 \\
   T(\omega) &=& \mid\! t(\omega)\!\mid^2 .
\end{eqnarray}\end{mathletters}

\subsection{Numerical Calculations}
\label{sec_the_nc}

The integral equation (\ref{eq_ietq}) is numerically solved by
converting it into a matrix equation through a quadrature scheme.
The details are given in the Appendix. It should be pointed out
that the discretization $q$-mesh is chosen in such a way that
$q=\pm k^R(\omega)$ are always points on the mesh, as required
by Eq. (\ref{eq_spprt}). In addition, the discretization is not
regular: the density of $q$-points around the poles at
$q=\pm k^R(\omega)$ is considerably larger ($\Delta q\approx
10^{-4}\omega/c$) than it is either in the radiative region
$|q|\leq\omega/c$ or in the non-radiative region away from the
poles ($\Delta q\approx 10^{-2}\omega/c$). The number $N$ of
$q$-points needed in the numerical procedure depends not only on
the accuracy required to sample the pole regions, but also on
the explicit form of the obstacle, which enters in the calculation
through its Fourier-transform in Eq. (\ref{eq_sq}). Throughout this
work, typically $N=2600$, except for the larger defects for which up to
$N=4000$ points are employed. The convergence of the numerical results
with increasing $N$ has been checked in the most unfavorable cases.

\section{RESULTS AND DISCUSSION}
\label{sec_res}

Note that up to now no restrictions have been imposed on the shape
of the obstacle apart from its having a finite extent along the
$x_1$-axis. Its surface impedance function $s(x_1)$ is connected to the
actual surface profile defined by $x_3 = f(x_1)$  through\cite{aamibc}
\begin{eqnarray}
    s(x) & = & -\frac{1-\epsilon(\omega)}{d(\omega)\epsilon(\omega)}
          [1-d^2(\omega)D^2]^{1/2} f(x_1) + O(f^2),
\label{eq_sh}
\end{eqnarray}
where $d(\omega)=(c/\omega)[-\epsilon(\omega)]^{-1/2}$ is the optical skin
depth, and $D\equiv d/dx_1$. In the case of small skin depths and surface
slopes $(dD)^2\ll 1$, the square root term on the rhs of Eq.
(\ref{eq_sh}) can be expanded as
\begin{eqnarray}
   \lefteqn{[1-(dD)^2]^{1/2}=  1-\frac{1}{2} (dD)^2-\ldots} \nonumber \\
   && \phantom{++}-\frac{1.1.3\ldots(2n-3)}{2.4.6\ldots 2n}(dD)^{2n}
   +O\bbox((dD)^{2n+2}\bbox).
\label{eq_ddn}
\end{eqnarray}
Then the Fourier transform of the surface impedance function, which is
needed in the calculation [cf. Eq. (\ref{eq_vq})], is related to the
Fourier transform $\hat{f}(Q)$ of the surface profile function through
\begin{eqnarray}
   \lefteqn{\hat{s}(Q)= -\frac{1-\epsilon(\omega)}{d(\omega)
     \epsilon(\omega)}\left(1-\frac{1}{2} [-i d(\omega)Q]^2\right.}
     && \nonumber \\ && \left.\phantom{++} -\frac{1}{8}[-i d(\omega)Q]^4
     +O\bbox([-i d(\omega)Q]^6\bbox)\right)\hat{f}(Q).
\label{eq_sqf}
\end{eqnarray}

In what follows, we will restrict the analysis to a Gaussian
defect of $1/e$ half-width $a$ and height $h$:
\begin{eqnarray}
    f(x_1) & = & h\exp\left(-x_1^2/a^2\right).
\end{eqnarray}
In addition, unless otherwise stated, we retain in Eq. (\ref{eq_ddn}),
and thus in Eq. (\ref{eq_sqf}), only the zeroth order term in the
expansion in powers of $[-i d(\omega)Q]^2$, as implicitly done in
Ref. \onlinecite{apl98}. Therefore the function $\hat{s}(Q)$ we will use
in our calculations is
\begin{mathletters}
\begin{eqnarray}
    \hat{s}(Q) &=& \pi^{1/2} s_0
        a\exp[-(aQ)^2/4],
\end{eqnarray}
with
\begin{eqnarray}
    s_0 &=& -\frac{1-\epsilon(\omega)}{\epsilon(\omega)}
            \frac{h}{d(\omega)}.
\end{eqnarray}
\label{eq_sqfg}
\end{mathletters}
It should be emphasized that the approximation involved in retaining only
the lowest order term in the expansion (\ref{eq_ddn}) affects only the
expression connecting the surface impedance with the real surface
profile, the scattering formulation being rigorous and energy conserving
(recall that losses are not accounted for) whatever the surface impedance
is. Nonetheless, inasmuch as we wish to be able to quantitatively relate
our results with real defect sizes, the effect of neglecting the higher
order terms in Eq. (\ref{eq_ddn}) has to be determined. We have thus
verified in the most unfavorable cases that including the first order
term in $[-id(\omega )Q]^2$ in Eq. (\ref{eq_sqf}) barely modifies our
calculations.

\begin{figure}
\epsfxsize=3in \epsfbox{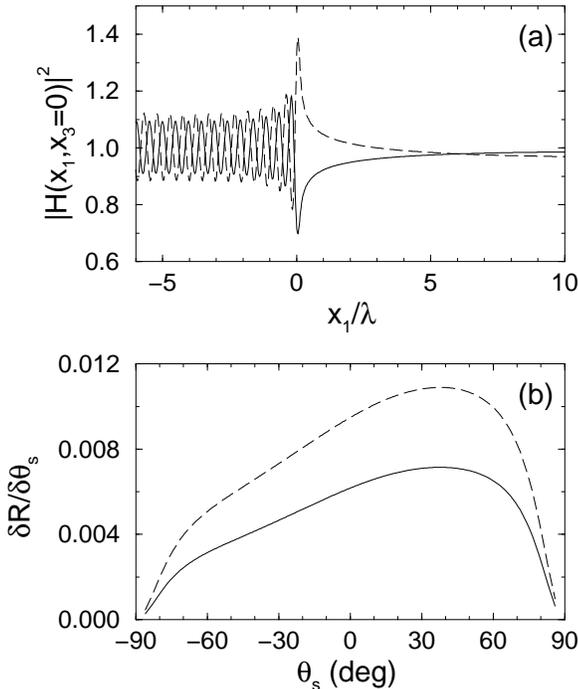}
\caption{(a) Square modulus of the total surface magnetic field and (b) DRC
         resulting from the scattering of a SPP of frequency
         $\hbar\omega=$1.96 eV ($\lambda=$632.8 nm) by Gaussian defects on a
         silver surface ($\varepsilon=-17.2$) of half-width $a=0.1\lambda$.
         Solid curve: $h=0.05\lambda$ (protuberance); dashed curve:
         $h=-0.05\lambda$ (indentation). }
\label{fig_hi}
\end{figure}

In order to establish the accuracy and efficiency of the numerical
calculations based on the formulation above, we first calculate the
function $T(q,\omega)$ [cf. Eq. (\ref{eq_ietq})], following the numerical
procedure outlined in Sec. \ref{sec_the} and the Appendix, for two
Gaussian defects of half-width $a/\lambda=0.1$ and heights
$h/\lambda=\pm 0.05$ (protuberance and indentation of equal height/depth),
where $\lambda$ is the wavelength of the SPP.  From these results, the SPP
reflection and transmission coefficients are straightforwardly calculated
[cf. Eqs. (\ref{eq_spprt}) and (\ref{eq_spprt2})], along with the DRC
[cf. Eqs. (\ref{eq_rtq}) and (\ref{eq_drc})]. Furthermore, the magnetic
and electric fields at any point in the vacuum half-space can be
calculated from Eqs. (\ref{eq_h}) and (\ref{eq_e}) by using Eq.
(\ref{eq_rtq}). In Fig.~\ref{fig_hi} we present the results thus obtained
for the magnetic field intensities at the vacuum-metal interface in the
vicinity of the Gaussian defects, and for the angular distribution of the
scattered field in the far field. From the surface magnetic field in
Fig.~\ref{fig_hi}(a), it is evident that both surface defects reflect
back part of the incoming SPP, which interferes with the incoming SPP
giving rise to the oscillatory pattern to the left of the defect
(negative $x_1$-axis). Near the defect the magnetic field is perturbed.
The outgoing transmitted SPP is seen to the right of the defect. The SPP
reflection and transmission coefficients are: $R_{SP}=0.0025$ and
$T_{SP}=0.9825$ for the protuberance, and $R_{SP}=0.0041$ and
$T_{SP}=0.9728$ for the indentation.
In Fig.~\ref{fig_hi}(b) a fairly uniform angular distribution of the DRC
is observed (this will be discussed below). The total scattered power
calculated
from Eq. (\ref{eq_s}) is $S=0.0149$ for the protuberance, and $S=0.0231$
for the indentation. Energy conservation is thus satisfied
within a $0.01\%$ error.

We find that away from the vicinity of the defect, the magnetic field is
fully described by either the interference between incoming and reflected
SPP on the left-hand side, or by merely the transmitted SPP on the
right-hand side. This corroborates, as expected, our argument in
Sec.~\ref{sec_the_rt} leading to Eqs. (\ref{eq_hxinf}). The reflection and
transmission coefficients are thus calculated from Eqs. (\ref{eq_spprt})
and (\ref{eq_spprt2}).

\subsection{Energy balance dependence on defect size}
\label{sec_res_rts}

The question now arises naturally as to how efficient the surface defect
is in coupling the incoming SPP into the different outgoing channels
(either SPP or photons), or conversely, what the appropriate defect
parameters are that maximize or minimize those channels; this is crucial
for both an understanding of the scattering process and the design of
practical devices. To that end, we have studied the dependence of the
scattering coefficients $R_{SP}$, $T_{SP}$,  and $S$ on the defect
half-width $a$ for both Gaussian protuberances and indentations of
different heights $h/\lambda=0.05$ and 0.2. The results are shown in
Fig.~\ref{fig_rts}. Several general features are evident from these
results.

First, SPP reflection is  relevant only for very narrow defects,
$a<\lambda/5$, for either protuberances or indentations. Indeed there
is an optimum defect width for which $R_{SP}$ is maximum.\cite{apl98}
These defects are called {\it plasmon mirrors}.\cite{smol,apl98} For
increasing defect widths, protuberances and indentations begin to
behave differently, except for their negligible contribution to SPP
reflection. On the one hand, SPP transmission through protuberances
monotonically diminishes at the expense of radiation. The conversion
is steeper the higher the defect is.
Indentations, however, exhibit an oscillatory pattern with increasing
defect width, in such a way that radiation (SPP transmission)
increases (decreases), passes through a maximum (minimum), and then
tends asymptotically to 0 (1). The oscillation period, the defect width
that yields maximum radiation, and the value of this maximum, all
depend on the surface height. Note that both protuberances and
indentation may behave as {\it light-emitters}\cite{apl98} (high
SPP-light conversion efficiency) for an appropriate (and distinct)
range of defect parameters. Below we analyze in detail the behavior of
SPP mirrors and light-emitters.

\subsection{Narrow defects: SPP mirrors}
\label{sec_res_mir}

Surface defects playing the role of SPP mirrors have been studied
experimentally in PSTM configurations.\cite{smol} This phenomenon has been
analyzed for different defect shapes in Ref. \onlinecite{apl98}, where in
addition a simple analytical prediction is given through a
perturbation-theoretic argument. In the case of Gaussian-shaped defects,
the predicted half-width that yields maximum reflection is:
$a_{mirr}\approx [2^{1/2}k^R(\omega)]^{-1}$. Our numerical
results further corroborate this prediction, since it is seen in
Fig.~\ref{fig_rts} that $a_{mirr}/\lambda\approx 0.1$ no matter what the
defect height is [as long as this height does not exceed the range of
validity of Eq. (\ref{eq_sh})]. Nonetheless, the maximum SPP reflection
increases with the defect height, and is slightly larger for indentations.

The electric and magnetic near-field intensities for Gaussian defects,
placed at the origin, of width $a/\lambda=0.1$ and heights $h/\lambda=\pm
0.05$ (protuberance and indentation, respectively) are presented in
Figs.~\ref{fig_nfpm_p} and~\ref{fig_nfpm_i}, respectively,. The near-field
maps are quite similar in both cases. The oscillations to the left of the
obstacles clearly reveal the interference  between the incident and
backscattered SPP, their period being $T\approx 2\pi/(2k^R)$, as expected,
and their contrast being related to $R_{SP}$. A bright region is seen to
the right of the defects that is due to the strong SPP transmission.
Poynting vector maps superimposed on the electric near-field intensity
maps confirm the description of the energy flow given above.

The corresponding angular distributions of scattered light (DRC) have been
shown in Fig.~\ref{fig_hi}(b). There are no significant qualitative
differences between protuberances and indentations, both yielding a fairly
structure-less angular dependence; quantitatively, an indentation leads to
stronger light coupling. The qualitative behavior is somewhat expected:
the same perturbation-theoretic argument predicting maximum SPP reflection
for a defect width that maximizes the scattering potential [cf. Eq.
(\ref{eq_vq})] at backscattering,\cite{apl98} leads to a mostly uniform
SPP coupling into EM waves in the radiative region ($|q|\leq\omega/c$).

\subsection{Wide protuberances: Total light-emitters}
\label{sec_res_wp}

The ability of Gaussian-shaped protuberances to couple SPP into light
has been pointed out in Ref. \onlinecite{apl98}. Here we analyze in detail
the conditions for protuberances and indentations alike to behave as
light-emitters with coupling efficiencies beyond 90$\%$, larger than that
reported in Ref. \onlinecite{apl98}. Figure~\ref{fig_rts} above illustrates
the discussion.

To begin with, let us focus on protuberances. For widths beyond those
producing significant SPP reflection (SPP mirrors), SPP-light conversion
increases monotonically whereas, as expected from energy conservation,
SPP transmission decays. This variation is faster for higher
protuberances. Indeed, the curves in Fig. \ref{fig_rts} (bottom) indicate
that $\lim_{a\rightarrow\infty}S=1$ even for the small protuberance.
We have found coupling efficiencies beyond 90$\%$ in the
case of $h/\lambda=0.2$ and $a/\lambda\geq 3.6$. In
Fig.~\ref{fig_nfle_p}, the electric and magnetic near-field
intensity maps for one such defect are shown. The absence of
oscillations to the left of the defect reveals that SPP reflection
is small; SPP transmission is small too (though considerably larger than
$R_{SP}$), as seen on the right-hand side of the defect.
A light beam is observed leaving the surface from the defect at
near-grazing scattering angles. This energy flow picture is further
corroborated by the angular distribution of the DRC shown in
Fig.~\ref{fig_ile}, with a maximum at $\theta_s\approx 74^{\circ}$.
Qualitatively, the fact that the metal protuberance enters  the vacuum
half-space seems to favor the SPP-photon coupling (playing the role of a
{\it launching platform}).

\subsection{Wide indentations: Light-emitters and SPP total transmission}
\label{sec_res_wi}

\begin{figure}
\epsfxsize=3in \epsfbox{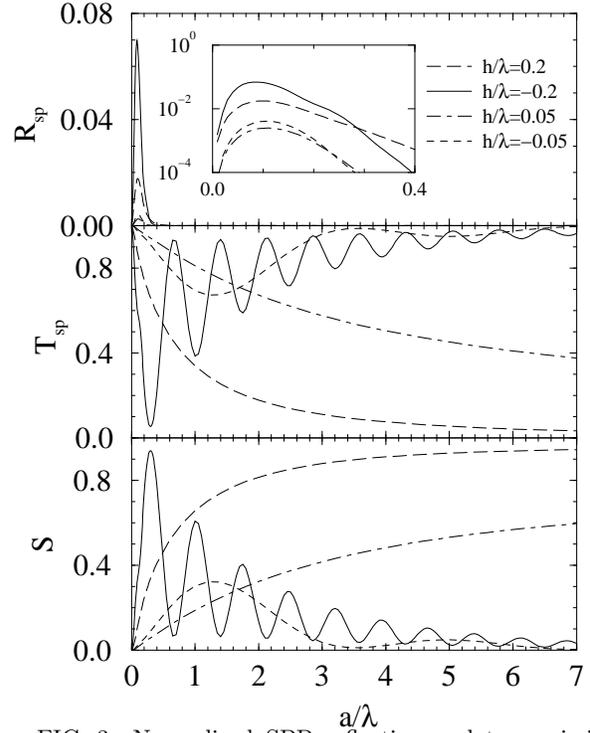}
\caption{Normalized SPP reflection and transmission coefficients ($R_{SP}$
         and $T_{SP}$, respectively), and total radiated energy S, as
         functions of the Gaussian defect half-width: $\hbar\omega=$1.96 eV
         ($\lambda=$632.8 nm) and $\varepsilon=-17.2$. Long-dashed curve:
         $h=0.2\lambda$; solid curve: $h=-0.2\lambda$; dot-dashed curve:
         $h=0.05\lambda$; dashed curve: $h=-0.05\lambda$. The inset zooms
         in the reflection coefficient in a semi-log scale for narrow
         defects. }
\label{fig_rts}
\end{figure}

\begin{figure}
\epsfxsize=3in \epsfbox{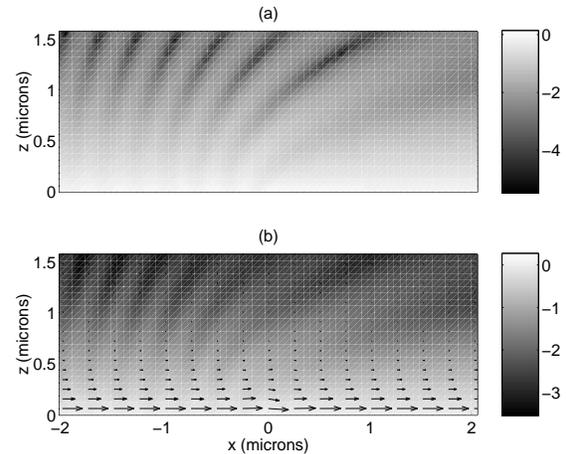}
\caption{Near field intensity distribution (in a log scale) resulting from
         the scattering of a SPP of frequency $\hbar\omega=$1.96 eV
         ($\lambda=$632.8 nm) by a Gaussian protuberance on a silver surface
         ($\varepsilon=-17.2$) of half-width $a=0.1\lambda$ and height
         $h=0.05\lambda$. (a) Magnetic field intensity and (b) Electric
         field intensity and Poynting vector.}
\label{fig_nfpm_p}
\end{figure}

\begin{figure}
\epsfxsize=3in \epsfbox{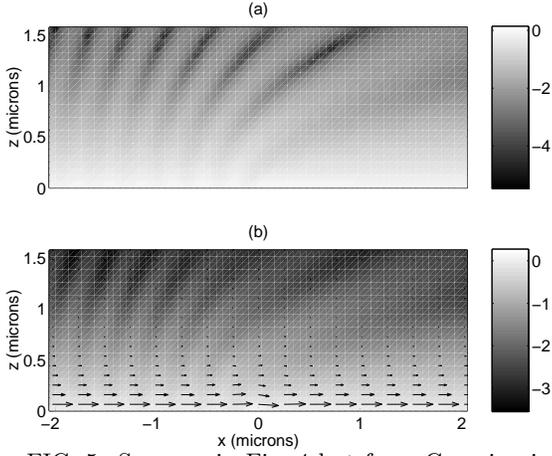}
\caption{Same as in Fig.~\protect{\ref{fig_nfpm_p}} but for a Gaussian
         indentation $h=-0.05\lambda$.}
\label{fig_nfpm_i}
\end{figure}

\begin{figure}
\epsfxsize=3in \epsfbox{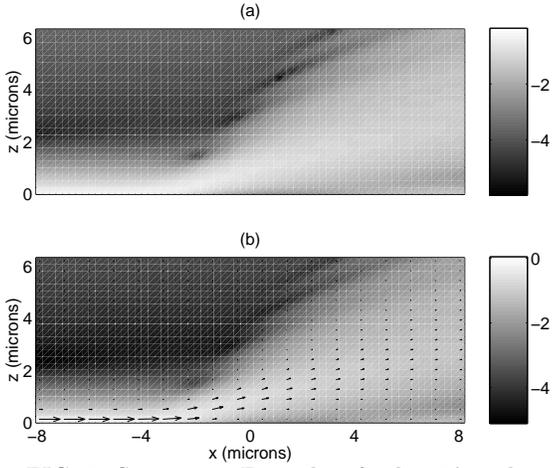}
\caption{Same as in Fig.~\protect{\ref{fig_nfpm_p}} but for $h=2\lambda$
         and $a=4\lambda$ (and for a larger near field area).}
\label{fig_nfle_p}
\end{figure}

\begin{figure}
\epsfxsize=3in \epsfbox{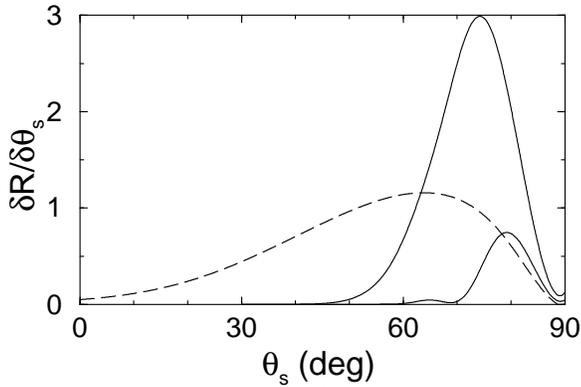}
\caption{DRC as in Fig.~\protect{\ref{fig_hi}}(b) but for: Upper solid curve,
         $h=2\lambda$ and $a=4\lambda$; dashed curve, $h=-2\lambda$ and
         $a=0.3\lambda$; lower solid curve, $h=-2\lambda$ and $a=4\lambda$.}
\label{fig_ile}
\end{figure}

In the case of wide indentations, however, the behavior of the
different outgoing channels differs from that for protuberances, and
exhibits a richer phenomenology. Upon increasing the width of the
indentation beyond the range of significant SPP reflection
(see Fig.~\ref{fig_rts}), SPP
transmission reaches a minimum value leading to maximum radiation, and
then slowly grows towards {\it total transmission} (no radiation) in
an oscillatory manner. The defect width that yields maximum radiation,
its value, and the oscillations depend on the defect height.

\begin{figure}
\epsfxsize=3in \epsfbox{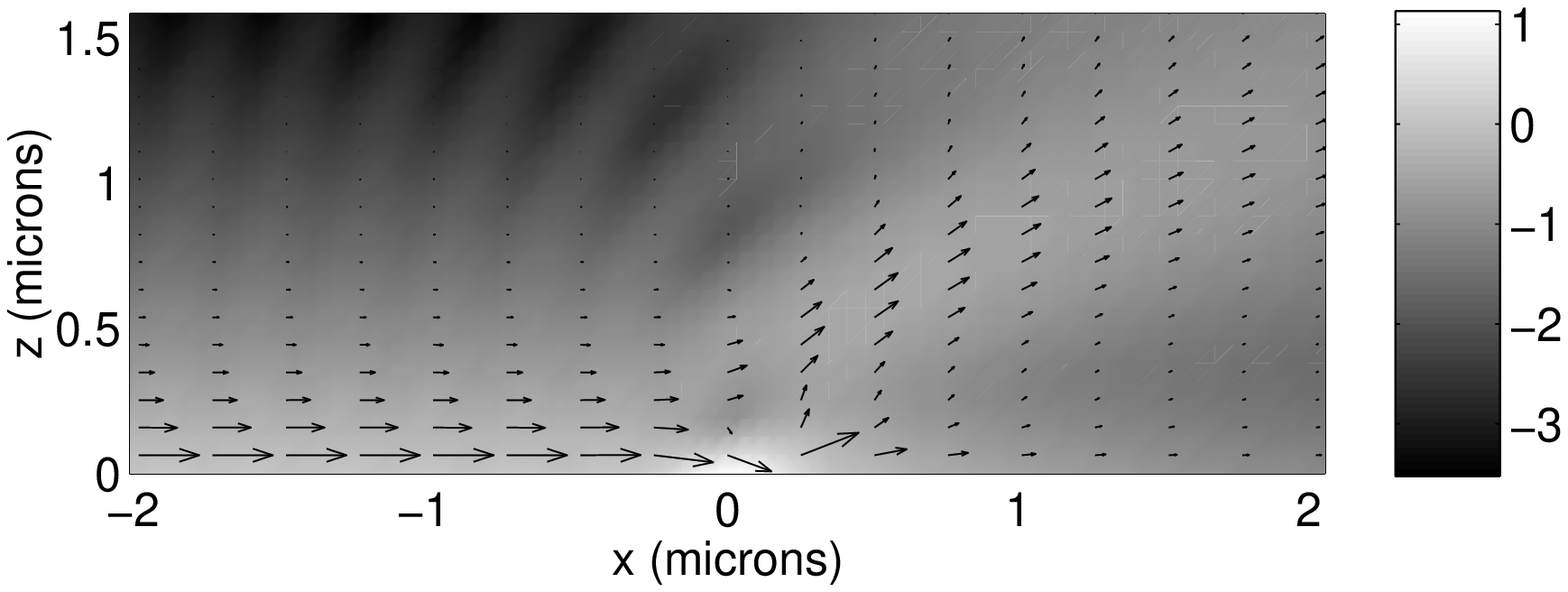}
\caption{Same as in Fig.~\protect{\ref{fig_nfpm_p}}(b) but for $h=-2\lambda$
         and $a=0.3\lambda$.}
\label{fig_nfle_i_3}
\end{figure}

\begin{figure}
\epsfxsize=3in \epsfbox{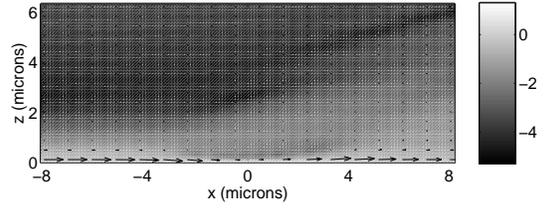}
\caption{Same as in Fig.~\protect{\ref{fig_nfle_p}}(b) but for $h=-2\lambda$
         and $a=4\lambda$.}
\label{fig_nfle_i4}
\end{figure}

To understand such behavior, we plot in Figs.~\ref{fig_nfle_i_3}
and~\ref{fig_nfle_i4} the electric near-field intensity maps for the
higher defects ($h/\lambda=0.2$) of widths $a/\lambda=0.3$ and
$a/\lambda=4$, respectively. These widths correspond to the first
absolute and sixth subsidiary, maxima in Fig.~\ref{fig_rts} (bottom),
respectively.  Both indentations give rise to
a negligible amount of SPP reflection (no oscillations to the left of
the defect), as expected from Fig.~\ref{fig_rts} (top). SPP transmission
(to the right of the defect) is very small for $a/\lambda=0.3$, but
a strong light beam at grazing scattering is observed ($S=94\%$). For
$a/\lambda=4$, although most of the energy goes into $T_{SP}=86.1\%$,
a small amount of radiation also at grazing scattering angles is seen
[recall that even though for this width a local maximum occurs in $S$,
its value is very small $S=13.3\%$, see Fig~\ref{fig_rts} (bottom)].
But what is very illustrative to the discussion on the behavior of the
outgoing channels is the near-field within the indentation (strictly
speaking, right on top of the indentation region, since we are using an
IBC on a flat surface). Oscillations are found therein, the number of
minima (one in Fig.~\ref{fig_nfle_i_3} and six in
Fig.~\ref{fig_nfle_i4}) being directly related to the position of
the corresponding maxima in the $S$ vs. $a$ curve [see
Fig.~\ref{fig_rts}]. Therefore, it can be inferred that the
oscillatory behavior of SPP transmission and conversion into light in
indentations is governed by a cavity-like effect.
In fact, the near field map (not shown here) in the
vicinity of any minimum-radiation indentation provides further evidence
for this suggestion.

As a consequence, the range of defect widths for which high coupling
efficiencies are encountered is far more restrictive for
indentations. In fact, in contrast with protuberances, only sufficiently
deep indentations ($h/\lambda\geq 0.2$) are capable of
producing radiation efficiencies $S>90\%$, and only for a narrow range
of parameters. It seems as if the indentation geometry would somehow
hinder grazing light scattering.
Therefore, this kind of {\it light-emitters} might not correlate with
any of the reciprocal versions (SPP flashlights) seen in PSTM
experiments.\cite{smol,bozh}

\subsection{Large width limit}
\label{sec_res_lwl}

Although the energy conservation criterion is reasonably well
satisfied in our calculations, even for defects wider than those used in
Fig.~\ref{fig_rts} (we have reached up to $a/\lambda=20$), one has to be
careful when interpreting the results in the limit
$a/\lambda\rightarrow\infty$.
It turns out that the determination
of the behavior of defects in this limit is important, since different
tendencies have been encountered for protuberances and indentations
(total radiation and transmission, respectively).

The analysis of the appropriate defect width that yields maximum coupling
is not simple even if making use of the Born approximation, since it
requires the evaluation of the integral of $\hat{s}(q-k^R)$ for all
homogeneous waves $|q|<\omega/c$. And yet such an  approximation does not
properly describe the formally exact numerical calculations.
Alternatively, we have carried out an
analytic calculation based on the use of a boundary condition similar to
the Kirchhoff approximation. The approach relies on the expression for
the scattering amplitude in terms of an integral equation along the
surface with the magnetic field and its normal derivative inside the
integrand (cf. Refs. \onlinecite{josaa91,ann} for the integral equation
formulation, and Ref. \onlinecite{prb96} for its version making use of
the IBC on a flat surface). By assuming that the surface magnetic field
is given by the incoming SPP, the scattering amplitude reads in the large
width limit:
\begin{eqnarray}
 \lim_{a/\lambda\rightarrow\infty} R(q,\omega)=
  2\pi\delta\bbox(q-k^R(\omega)\bbox)(1+\pi s_0).
\end{eqnarray}
Although it does not satisfy energy conservation (not surprisingly, due to
the approximation involved), the former result gives an estimation of the
limiting behavior shown above (see Fig.~\ref{fig_rts}): SPP transmission
saturates for indentations ($s_0>0$), whereas protuberances ($s_0<0$)
tend to decrease SPP transmission (in agreement with the numerical
calculations, yet the exact limit is not predicted).

\section{CONCLUSIONS}
\label{sec_con}

We have presented a theoretical formulation that describes in a
rigorous manner the scattering of a surface plasmon polariton
propagating along a planar vacuum-metal interface by a
one-dimensional obstacle modeled through an impedance boundary
condition. By solving the $k$-space scattering integral equation
upon which the formulation is based, the angular spectrum of
the scattered electromagnetic field in the vacuum half-space
above the metal surface can be calculated, which in turn allows
us to obtain the near electric and magnetic fields, the amplitudes
of the reflected and transmitted SPP, and the angular distribution
of the intensity of radiated waves resulting from the conversion of
SPP into volume waves. A numerical method to solve the scattering
integral equation has been put forth.

We have made use of these calculation methods to study the SPP
scattering by one-dimensional Gaussian defects, either protuberances
or indentations. In particular, the dependence of the scattering
process on the surface defect parameters has been analyzed.
Several conclusions can be drawn from our results with respect to the
behavior of Gaussian protuberances or indentations.

SPP reflection is only significant for very narrow surface defects,
with half-widths $a<c/\omega$. Our near field results explicitly show
that in this case protuberances and indentations behave alike, the
latter reflecting SPP slightly more efficiently.
The dependence of the SPP reflection coefficient on the half-width
confirms for different defect heights the condition predicted in Ref.
\onlinecite{apl98} of maximum SPP reflection, leading to the {\it
plasmon mirrors} seen in PSTM experiments.\cite{smol}

For wider Gaussian defects, protuberances and indentations yield a
entirely different picture, the only common feature being the negligible
contribution to SPP reflection. Protuberances, on the one hand,
increasingly radiate more light at near grazing scattering angles at the
expense of SPP transmission. They behave as {\it light-emitters} with
coupling efficiencies approaching 100\% with increasing half-width. The
higher the defect is, the larger the SPP-light conversion.
On the other hand, indentations tend to total SPP transmission without
radiation with increasing half-width. The increase (decrease) of the SPP
transmission (light coupling) occurs in an oscillatory manner starting
from an absolute minimum (maximum) in transmission (radiation) for small
half-widths, the period of the oscillations being related to the defect
impedance in a way reminiscent of a cavity-like effect. Interestingly, we
have found that for sufficiently deep indentations, this maximum
radiation value can be extremely large (even larger than 90\%), so that
the Gaussian indentation thus behave as a light-emitter.

Our results and discussion provide a thorough picture of the different
aspects of SPP  scattering by surface defects which, besides being
interesting in itself as a scattering process, appears to be useful in a
number of related problems.
\cite{ebbe98,schro98,nie97,klar98,smol95,smol,bozh,smol99,pasc98,apl98}
It is rigorous for one-dimensional defects and indeed sheds light on the
two-dimensional case, and can in turn explain and predict experimental
results.\cite{smol,bozh} In this regard, it would be interesting to
perform experiments on metal surfaces with defects of controled profile.
With respect to the 2D case, it should be emphasized that the recent work
by Shchegrov {\it et al.}\cite{snm97} for circularly symmetric surface
defects reproduces the radiation pattern with peculiar lobes in the
azimuthal angle dependence observed experimentally.\cite{smol}
However, further theoretical work is needed that could
address more complicated geometries used in the experiments and/or
unexplained processes involving surface plasmon-polaritons.
\cite{ebbe98,nie97,smol99}

\acknowledgments

This work was supported in part by Army Research Office Grant No.
DAAH 0-96-1-0187, and by both the Spanish DGES (Grants No.
PB96-0844-C02-02 and PB97-1221) and the Consejo Superior de
Investigaciones Cient{\'\i}ficas.
J.A.S.-G. acknowledges fruitful discussions with J. M. S{\'a}iz.

\appendix
\section*{}

By discretizing $q$ and $p$ in Eq. (\ref{eq_ietq}) and replacing  the
infinite limits in the integral by sufficiently large finite limits, the
following system of linear equations is obtained for the
$t_n\equiv T(p_n,\omega)$ unknowns:
\begin{eqnarray}
    [K_{mn}]t_n &=& v_m ,
\end{eqnarray}
with $v_m\equiv V\bbox(q_m|k^R(\omega)\bbox)$. The matrix elements
$K_{mn}$ are given by
\begin{eqnarray}
    K_{mn} &=& \delta_{mn}+M_{mn} ,
\end{eqnarray}
where $\delta_{mn}$ is the Kronecker delta and
\begin{mathletters}\begin{eqnarray}
    M_{mn}= & & -\frac{1}{2\pi} C\bbox(\pm k^R(\omega),\omega\bbox)
      V\bbox(q_m|\pm k^R(\omega)\bbox) \nonumber \\ &&
      \times{\displaystyle \left(\pi i+\log\left|\frac{2k^R(\omega)-
      \Delta q/2}{2k^R(\omega)+\Delta q/2}\right|\right),} \nonumber \\
      && \phantom{+++++++++} p_n=\pm k^R(\omega) \\ = & &
      -\frac{1}{2\pi} C(q_m,\omega) V(q_m|p_n) \nonumber \\ && \times
      {\displaystyle \left(\log\left|
      \frac{p_n+\Delta q/2-k^R(\omega)}{p_n-\Delta q/2-k^R(\omega)}\right|
      \right. } \nonumber \\ && + {\displaystyle \left.\log\left|
      \frac{p_n-\Delta q/2+k^R(\omega)}{p_n+\Delta q/2+k^R(\omega)}\right|
      \right),} \nonumber \\ && \phantom{+++++++++} p_n\neq\pm k^R(\omega).
\end{eqnarray}\end{mathletters}
In obtaining Eqs. (A3), the explicit form of $G_0(q,\omega)$ shown in
Eq. (\ref{eq_gqri}) has been taken into account in calculating the
corresponding integrals over the sampling intervals
$[p_n-\Delta q/2,p_n+\Delta q/2]$.

\begin{references}
\bibitem{cauchy} In so doing, we are assuming that $T(q,\omega)$ as a
             function of complex $q$ is analytical and behaves
             appropriately in the limit $|q|\rightarrow\infty$.
\end{references}
\end{document}